\newcommand{\extended}[1]{}    
\newcommand{\short}[1]{#1}

\documentclass{article}
\usepackage{ijcai23}

\usepackage{times}
\usepackage{soul}
\usepackage{url}
\usepackage[hidelinks]{hyperref}
\usepackage[utf8]{inputenc}
\usepackage[small]{caption}
\usepackage{graphicx}
\usepackage{amsmath}
\usepackage{amsthm}
\usepackage{booktabs}
\usepackage[switch]{lineno}

\usepackage{adjustbox}
\usepackage{fancyvrb}
\usepackage{dcolumn}
\usepackage{amssymb}

\usepackage{multirow} \usepackage{hhline}   \usepackage{subcaption}

\newcolumntype{d}[1]{D{.}{.}{#1} }

\newcommand{\lan}[1]{\ensuremath{\mathbf{#1}}\xspace}

\newcommand{\stratstyle}[1]{\ensuremath{\mathrm{#1}}}

\newcommand{\ATL}[1][]{\lan{ATL_{\stratstyle{#1}}}}

\newcommand{\atl}[1][]{\ATL[#1]}

\newcommand{\CSL}[1][]{\lan{CSL}}
\newcommand{\CSLP}[1][]{\lan{CSLP}}

\newcommand{\MIATL}[1][]{IATL[R]}

\newcommand{\ATLES}[1][]{\lan{ATLES}}
\newcommand{\ATELA}[1][]{\lan{ATELA}}

\newcommand{\ACTL}{\lan{ACTL}}

\newcommand{\CTLs}{\lan{CTL^\star}}

\newcommand{\coop}[2][]{\langle\!\langle{#2}\rangle\!\rangle_{_{\!\mathit{#1}}}}

\newcommand{\Epath}{\mathsf{E}}
\newcommand{\Apath}{\mathsf{A}}

\newcommand{\TransArrow}[1][]{\hookrightarrow....}

                                    \newcommand{\onepath}[1][]{\ensuremath{\lambda\ifthenelse{\equal{#1}{}}{}{[#1]}}}

\newcommand{\plaus}[1][]{\ifthenelse{\equal{#1}{}}{\mathbf{P\;\!\!l}\,}{\mathbf{P\;\!\!l}_{#1}\,}}
\newcommand{\phys}[1][]{\ifthenelse{\equal{#1}{}}{\mathbf{P\;\!\!h}\,}{\mathbf{P\;\!\!h}_{#1}\,}}

\newcommand{\plaumodels}[1][]{\ensuremath{\ifthenelse{\equal{#1}{}}{\models_\sPlaupaths}{\models_{#1}}}}
\newcommand{\sPlaupaths}{\ensuremath{P}}

\newcommand{\then}{\rightarrow}

\usepackage{color}

\definecolor{lightgrey}{rgb}{0.8,0.8,0.8}
\definecolor{grey}{rgb}{0.6,0.6,0.6}
\definecolor{darkgrey}{rgb}{0.4,0.4,0.4}
\definecolor{darkgreen}{rgb}{0,0.7,0}

\newcommand{\shift}{3pt}

\newcommand{\set}[1]{\{{#1}\}}

\newcommand{\complexityclass}[1]{\ensuremath{\mathbf{{#1}}}\xspace}

\newcommand{\Pspace}{\complexityclass{PSPACE}}

\newcommand{\PSPACE}{\Pspace}

\newcommand{\putaway}[1]{}

\newcommand{\para}[1]{\smallskip\noindent\textbf{#1}}

\newenvironment{itemize2}{\begin{itemize}[noitemsep]}{\end{itemize}}

\newenvironment{description2}{\begin{description}[noitemsep]}{\end{description}}

\newcommand{\finis}{{\scriptsize $\blacksquare$}}
\newcommand{\finisdef}{$\Box$}
\newcommand{\bul}{{\tiny $\blacksquare$}}

\def\itemiremember{\labelitemi}
\def\itemiiremember{\labelitemii}

﻿
\RequirePackage{amsthm}
\RequirePackage{amsmath} \RequirePackage{ifxetex,ifluatex}
\RequirePackage{latexsym}
\RequirePackage{enumitem} \setitemize{noitemsep,parsep=0pt,partopsep=0pt,leftmargin=15pt}
\setenumerate{noitemsep,parsep=0pt,partopsep=0pt,leftmargin=20pt}
   \RequirePackage{varioref}
   \RequirePackage{hyperref}
   \RequirePackage[capitalize]{cleveref}
\RequirePackage{xcolor}
\RequirePackage{tikz}
\RequirePackage{subcaption} 
\RequirePackage{graphicx}
\RequirePackage{thmtools}

\RequirePackage{fancyvrb}
\RequirePackage{calc} \RequirePackage[euler]{textgreek}
\usepackage[ruled]{algorithm2e}

\usepackage[normalem]{ulem} 
\SetKw{Continue}{continue}
    \LinesNumbered
    \let\oldnl\nl \newcommand{\nonl}{\renewcommand{\nl}{\let\nl\oldnl}} \SetKwFunction{approxLocalDomain}{ApproxLocalDomain}
	\SetKwProg{myalg}{Algorithm}{}{}
	\SetKwProg{myproc}{}{}{}
	\SetKwInOut{Input}{input}
	\SetKwInOut{Output}{output}
	\SetKwFunction{procEdgeLabels}{ProcEdge}
    \SetKwFunction{procPreCond}{ProcPreCond}
    \SetKwFunction{enqueueSucc}{EnqueueDirectSucc}
    \SetKwFunction{procSelfLoops}{ProcSelfLoops}
    \SetKwFunction{procIncEdges}{ProcIncEdges}
	\SetKwFunction{computeAbstraction}{ComputeAbstraction}
    \SetKwFunction{computeMergeAbstraction}{ComputeAbstraction}
\SetKwFunction{generalVarAbstraction}{ComputeAbstraction}
    \SetKwFunction{extractMax}{ExtractMax}
    \SetKwFunction{enqueue}{Enqueue}
	\SetAlgoLined\DontPrintSemicolon
\usepackage{microtype}

\makeatletter
\def\maxfitwidth{\ifdim\Gin@nat@width>\linewidth
	\linewidth
	\else
	\Gin@nat@width
	\fi
}
\makeatother

\newcommand\Var{\textit{Var}}

  \newcommand{\Uppaal}{\textsc{Uppaal}\xspace}

\newcommand{\sentencecase}[1]{{\expandafter\MakeUppercase#1}}

\newcommand\restr[2]{{\left.\kern-\nulldelimiterspace #1 \right\vert_{#2} }}

\RequirePackage{array} \newcolumntype{H}{>{\setbox0=\hbox\bgroup}c<{\egroup}@{}} %
\newdimen\proofrulebreadth \proofrulebreadth=.05em
\newdimen\proofdotseparation \proofdotseparation=1.25ex
\newdimen\proofrulebaseline \proofrulebaseline=2ex
\newcount\proofdotnumber \proofdotnumber=3
\let\then\relax
\def\hfi{\hskip0pt plus.0001fil}
\mathchardef\squigto="3A3B
\newif\ifinsideprooftree\insideprooftreefalse
\newif\ifonleftofproofrule\onleftofproofrulefalse
\newif\ifproofdots\proofdotsfalse
\newif\ifdoubleproof\doubleprooffalse
\let\wereinproofbit\relax
\newdimen\shortenproofleft
\newdimen\shortenproofright
\newdimen\proofbelowshift
\newbox\proofabove
\newbox\proofbelow
\newbox\proofrulename
\def\shiftproofbelow{\let\next\relax\afterassignment\setshiftproofbelow\dimen0 }
\def\shiftproofbelowneg{\def\next{\multiply\dimen0 by-1 }\afterassignment\setshiftproofbelow\dimen0 }
\def\setshiftproofbelow{\next\proofbelowshift=\dimen0 }
\def\setproofrulebreadth{\proofrulebreadth}

\def\prooftree{\ifnum  \lastpenalty=1
\then   \unpenalty
\else   \onleftofproofrulefalse
\fi
\ifonleftofproofrule
\else   \ifinsideprooftree
        \then   \hskip.5em plus1fil
        \fi
\fi
\bgroup \setbox\proofbelow=\hbox{}\setbox\proofrulename=\hbox{}\let\justifies\proofover\let\leadsto\proofoverdots\let\Justifies\proofoverdbl
\let\using\proofusing\let\[\prooftree
\ifinsideprooftree\let\]\endprooftree\fi
\proofdotsfalse\doubleprooffalse
\let\thickness\setproofrulebreadth
\let\shiftright\shiftproofbelow \let\shift\shiftproofbelow
\let\shiftleft\shiftproofbelowneg
\let\ifwasinsideprooftree\ifinsideprooftree
\insideprooftreetrue
\setbox\proofabove=\hbox\bgroup$\displaystyle \let\wereinproofbit\prooftree
\shortenproofleft=0pt \shortenproofright=0pt \proofbelowshift=0pt
\onleftofproofruletrue\penalty1
}

\def\eproofbit{\ifx    \wereinproofbit\prooftree
\then   \ifcase \lastpenalty
        \then   \shortenproofright=0pt  \or     \unpenalty\hfil         \or     \unpenalty\unskip       \else   \shortenproofright=0pt  \fi
\fi
\global\dimen0=\shortenproofleft
\global\dimen1=\shortenproofright
\global\dimen2=\proofrulebreadth
\global\dimen3=\proofbelowshift
\global\dimen4=\proofdotseparation
\global\count255=\proofdotnumber
$\egroup  \shortenproofleft=\dimen0
\shortenproofright=\dimen1
\proofrulebreadth=\dimen2
\proofbelowshift=\dimen3
\proofdotseparation=\dimen4
\proofdotnumber=\count255
}

\def\proofover{\eproofbit \setbox\proofbelow=\hbox\bgroup \let\wereinproofbit\proofover
$\displaystyle
}\def\proofoverdbl{\eproofbit \doubleprooftrue
\setbox\proofbelow=\hbox\bgroup \let\wereinproofbit\proofoverdbl
$\displaystyle
}\def\proofoverdots{\eproofbit \proofdotstrue
\setbox\proofbelow=\hbox\bgroup \let\wereinproofbit\proofoverdots
$\displaystyle
}\def\proofusing{\eproofbit \setbox\proofrulename=\hbox\bgroup \let\wereinproofbit\proofusing
\kern0.3em$
}

\def\endprooftree{\eproofbit \dimen5 =0pt\dimen0=\wd\proofabove \advance\dimen0-\shortenproofleft
\advance\dimen0-\shortenproofright
\dimen1=.5\dimen0 \advance\dimen1-.5\wd\proofbelow
\dimen4=\dimen1
\advance\dimen1\proofbelowshift \advance\dimen4-\proofbelowshift
\ifdim  \dimen1<0pt
\then   \advance\shortenproofleft\dimen1
        \advance\dimen0-\dimen1
        \dimen1=0pt
\ifdim  \shortenproofleft<0pt
        \then   \setbox\proofabove=\hbox{\kern-\shortenproofleft\unhbox\proofabove}\shortenproofleft=0pt
        \fi
\fi
\ifdim  \dimen4<0pt
\then   \advance\shortenproofright\dimen4
        \advance\dimen0-\dimen4
        \dimen4=0pt
\fi
\ifdim  \shortenproofright<\wd\proofrulename
\then   \shortenproofright=\wd\proofrulename
\fi
\dimen2=\shortenproofleft \advance\dimen2 by\dimen1
\dimen3=\shortenproofright\advance\dimen3 by\dimen4
\ifproofdots
\then
        \dimen6=\shortenproofleft \advance\dimen6 .5\dimen0
        \setbox1=\vbox to\proofdotseparation{\vss\hbox{$\cdot$}\vss}\setbox0=\hbox{\advance\dimen6-.5\wd1
                \kern\dimen6
                $\vcenter to\proofdotnumber\proofdotseparation
                        {\leaders\box1\vfill}$\unhbox\proofrulename}\else   \dimen6=\fontdimen22\the\textfont2 \dimen7=\dimen6
        \advance\dimen6by.5\proofrulebreadth
        \advance\dimen7by-.5\proofrulebreadth
        \setbox0=\hbox{\kern\shortenproofleft
                \ifdoubleproof
                \then   \hbox to\dimen0{$\mathsurround0pt\mathord=\mkern-6mu\cleaders\hbox{$\mkern-2mu=\mkern-2mu$}\hfill
                        \mkern-6mu\mathord=$}\else   \vrule height\dimen6 depth-\dimen7 width\dimen0
                \fi
                \unhbox\proofrulename}\ht0=\dimen6 \dp0=-\dimen7
\fi
\let\doll\relax
\ifwasinsideprooftree
\then   \let\VBOX\vbox
\else   \ifmmode\else$\let\doll=$\fi
        \let\VBOX\vcenter
\fi
\VBOX   {\baselineskip\proofrulebaseline \lineskip.2ex
        \expandafter\lineskiplimit\ifproofdots0ex\else-0.6ex\fi
        \hbox   spread\dimen5   {\hfi\unhbox\proofabove\hfi}\hbox{\box0}\hbox   {\kern\dimen2 \box\proofbelow}}\doll \global\dimen2=\dimen2
\global\dimen3=\dimen3
\egroup \ifonleftofproofrule
\then   \shortenproofleft=\dimen2
\fi
\shortenproofright=\dimen3
\onleftofproofrulefalse
\ifinsideprooftree
\then   \hskip.5em plus 1fil \penalty2
\fi
}

\definecolor{tucgreen}{RGB}{0,140,79}
\newcommand{\YK}[1]{{\color{blue}#1}}
\newcommand{\todo}[1]{{\color{red}#1}}

\newcommand{\NCand}{\ensuremath{\mathit{NC}}}
\newcommand{\NVot}{\ensuremath{\mathit{NV}}}

  \newcommand{\worklaptopcpu}{Intel i7-8665U 2.11 GHz CPU}

\newtheorem{example}{Example}

\title{Practical Model Reductions for Verification of Multi-Agent Systems}

\author{Wojciech Jamroga$^{1,2}$
\and
Yan Kim$^1$
\affiliations
$^1$Interdisciplinary Centre for Security, Reliability, and Trust, SnT, University of Luxembourg\\
$^2$Institute of Computer Science, Polish Academy of Science, Warsaw, Poland
\emails
\{wojciech.jamroga, yan.kim\}@uni.lu
}

\begin{document}

\maketitle

\begin{abstract}
Formal verification of intelligent agents is often computationally infeasible due to state-space explosion.
We present a tool for reducing the impact of the explosion by means of state abstraction that is (a) easy to use and understand by non-experts, and (b) agent-based in the sense that it operates on a modular representation of the system, rather than on its huge explicit state model.
\end{abstract}

\section{Introduction}

\emph{Multi-agent systems (MAS)}~\cite{Wooldridge02intromas,Shoham09MAS} describe interactions of autonomous {agents}, often assumed to be intelligent and/or rational. With the development of Internet and social networks, the impact of MAS on everyday life is becoming more and more significant. At the same time, their complexity is rapidly increasing. In consequence, formal methods for analysis and verification of MAS are badly needed.

\para{Verification and model reduction.}
Algorithms and tools for verification have been in constant development for 40 years, with temporal model checking being most popular~\cite{Baier08mcheck,Clarke18principles}.
The main obstacle for \emph{practical} use of those techniques is state-space explosion. Model checking of MAS with respect to their \emph{modular representations} ranges from \PSPACE-complete to undecidable~\cite{Schnoebelen03complexity,Jamroga15specificationMAS}.
A possible way to mitigate the complexity is by model reductions, such as abstraction refinement~\cite{Clarke00cegar} and partial-order reduction~\cite{Peled93representatives}. Unfortunately, lossless reductions (i.e., ones that produce fully equivalent models) are usually too weak, in the sense that the resulting model is still too large for feasible verification.

\para{Towards practical abstraction.}
In this work, we revisit the idea of lossy state abstraction~\cite{Cousot77abstraction,Clarke94abstraction}, and in particular \emph{may/must abstraction}~\cite{Godefroid01abstractionbased} that potentially removes relevant information about the system, but produces arbitrarily small reduced models. Such verification works best with users who are knowledgeable about the application domain, as its conclusiveness crucially depends on what aspects of the model are being removed. Ideally, the user should be a domain expert, which often implies no in-depth knowledge of verification algorithms. This calls for a technique that is easy to use and understand, preferably supported by a Graphical User Interface (GUI). Moreover, the abstraction should be \emph{agent-based} in the sense that it operates on modular representations of the MAS, and does not require to generate the full explicit-state model before the reduction. The theoretical backbone of our abstraction scheme is presented in~\cite{jamroga2022practical}.
Here, we report on the implementation, and show its usefulness through case studies.

\para{Contribution.}
We propose a tool for reduction of MAS models by removing an arbitrary subset of variables from the model specification. After the user selects the variables to be removed, the tool can produce two new model specifications: one guaranteed to overapproximate, and one to underapproximate the original model. Then, the user can verify properties of the original model by model checking the new specifications with a suitable model checker.
Our model specifications are in the form of \emph{MAS Graphs}~\cite{jamroga2022practical}, a variant of automata networks with asynchronous execution semantics and synchronization on joint action labels~\cite{Priese83apa-nets,Jamroga20POR-JAIR}.
As the model checker of choice, we use \Uppaal~\cite{Behrmann04uppaal-tutorial}, one of the few temporal model checkers with GUI.

Our tool provides a simple command-line interface, where the user selects the input file with a model specification prepared in \Uppaal, the variables to be abstracted away, and the abstraction parameters. It outputs a file with the over- (resp.~under-)approximating model specification, that can be opened in \Uppaal for scrutiny and verification.
The source code and examples are available at \texttt{\url{https://tinyurl.com/ijcai-demo}}.
Importantly, the abstraction uses modular representations for input and output; in fact, it does \emph{not} involve the generation of the global state space at all.
To our best knowledge, this is the first tool for practical user-defined model reductions in model checking of MAS.

\para{Related work.}
The existing implementations of state abstraction for temporal model checking concern mostly automated abstraction. In particular, \short{CEGAR~}\extended{Counterexample-Guided Abstraction Refinement (CEGAR)~}\cite{Clarke00cegar,Clarke03cegar} has been implemented for NuSMV~\cite{Cimatti02nusmv}, and 3-valued abstraction~\cite{Godefroid01abstractionbased,godefroid2014may} was implemented in Yasm~\cite{Gurfinkel06yasm} and YOGI~\cite{Godefroid10yogi}. In each case, abstraction involves the generation of the global state space, which is the main bottleneck when verifying MAS.
Other, user-defined abstraction schemes have been defined only theoretically~\cite{Shoham04abstraction,Ball06abstraction,Dams18abstraction+refinement}, and also require to generate all global states and/or transitions.
The approaches in~\cite{cohen2009abstraction,Belardinelli19abstractionStrat} come closest to our work, as they use modular representations of the state space. However, they both need a global representation of the transition space, and no implementation is reported.

\section{Formal Background}

\begin{figure}[t]
    \includegraphics[width=\columnwidth]{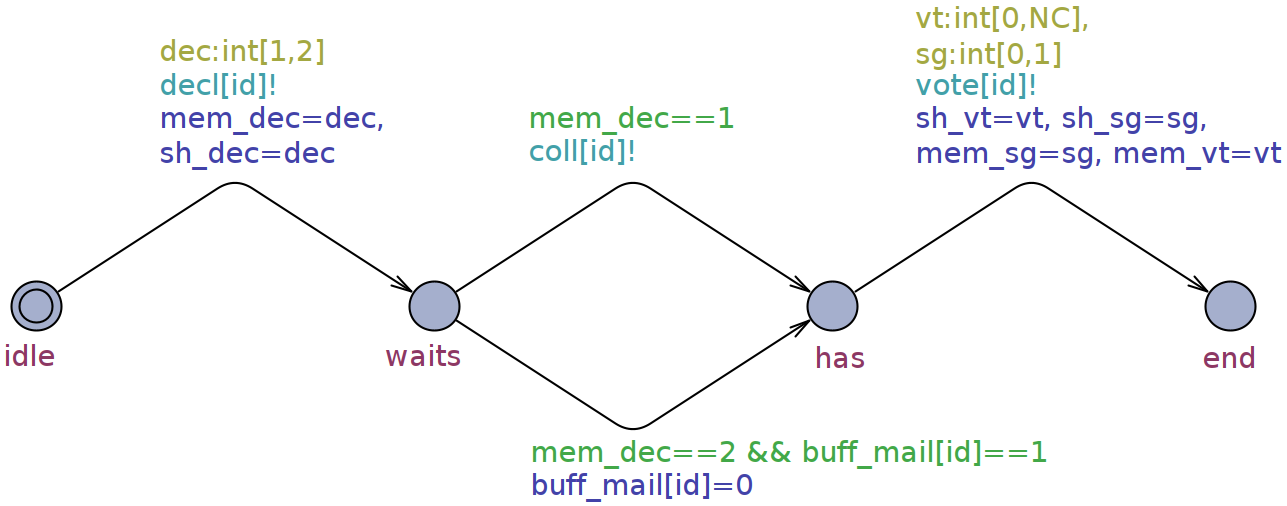}
    \caption{\emph{Voter} template.
    The agent first declares if she prefers to receive the election package by post (\texttt{dec=2}) or in person (\texttt{dec=1}). Then, she waits until it can be collected, and casts the ballot together with her voting card.
    The \emph{select} label for edge \texttt{idle}$\xrightarrow{}$\texttt{waits} (resp.~\texttt{has}$\xrightarrow{}$\texttt{end}) specifies a nondeterministic choice of the value
of variable $\texttt{dec}\in\set{1,2}$ (resp.~$\texttt{vt}\in\set{1,\ldots,NC}$ and $\texttt{sg}\in\set{0,1}$)
    }
    \label{fig:voter}
\end{figure}

\para{MAS graphs and templates.}
To specify the system to be verified, we use \emph{MAS graphs}\extended{~\cite{jamroga2022practical}}, based on standard models of concurrency\extended{~\cite{Priese83apa-nets}}, and compatible with \Uppaal model specifications\extended{~\cite{Behrmann04uppaal-tutorial}}.
A \emph{MAS graph} is a multiset of \emph{agent graphs}, possibly sharing a set of \emph{global variables}.
Each agent graph includes finitely many \emph{locations} and \emph{private variables} that, together, define its local state space.
Moreover, \emph{edges} between locations determine the local transition relation.
Each edge can be labelled with
a randomized \emph{selection} command\extended{ (a pair of variable and range, from which it can bound to a value),\footnote{
  Note that an edge with select label is equivalent to a set of almost identical edges with selector occurrences substituted with actual values taken from a given range. }
},
boolean \emph{precondition}\extended{ (condition over variables, that must be satisfied for the edge to be taken)},
\emph{synchronisation} command\extended{ (a channel name followed by `!' for sending or `?' for receiving)},
and/or a \emph{postcondition} updating the values of some variables.
A synchronizing edge can only be taken with a complementary one in another agent. \extended{
  We only consider systems with variables ranging over a finite domain of integers and.\footnote{Uppaal v4.1.24 supports 16-bit integers, optionally bounded.}
}An example agent graph is shown in~\autoref{fig:voter}.

A \emph{MAS template} treats each agent graph as a template, and specifies the number of its instances that occur in the verified system (each differing only by the value of variable \texttt{id}).

\para{Models.}
Every MAS graph $G$ can be transformed to its \emph{combined MAS graph}: technically, a single agent graph $comb(G)$ given by the asynchronous product of the agent graphs in $G$. Each location in $comb(G)$ is a tuple of agents' locations in $G$. Moreover, the set of variables in $comb(G)$ is the union of all variables occurring in $G$.
A \emph{global model} is obtained from $comb(G)$ by unfolding it to the labelled transition system where states are defined by combined locations and valuations of all the variables.
Such models are usually huge, and create an important bottleneck in model checking MAS.

\para{Formal verification and model reduction.}
Our tool addresses model checking of temporal properties expressed in the well known branching-time logic \CTLs~\cite{Emerson90temporal}\extended{, cf.~\cite{Baier08mcheck} for a textbook on temporal model checking}.
To mitigate the impact of state-space explosion, we use \emph{state abstraction}, i.e., a method that reduces the state space by clustering similar \emph{concrete states} into a single \emph{abstract state}. In order for the scheme to be practical, it must be easy to use, and avoid the generation of the concrete global model.
We summarize the details of our abstraction scheme in the next section.

\section{Abstraction by Removal of Variables}

\extended{
  The simplest way to reduce a MAS graph is to remove some model variables, or merge them to a new variable containing less information than the original ones.
}
Our tool employs the abstraction scheme of~\cite{jamroga2022practical}, and
produces specifications of two abstract models: a \emph{may-abstraction} (that overapproximates the concrete states and transitions) and a \emph{must-abstraction} (that underapproximates them).
Consequently, if a universal \CTLs formula is true in the \emph{may}-abstraction, then it must be true in the concrete model, and if it is false in the \emph{must}-abstraction, then it must be false in the concrete model\extended{~\cite{jamroga2022practical}}.

\para{Variable removal.}
In the simplest variant, the abstraction concerns a complete removal of some variables $V\subseteq\Var$ from the model specification.
For example, one might remove variables \texttt{mem\_vt}, \texttt{mem\_sg} from the agent graph in~\autoref{fig:voter}, i.e., the voter's memory of the cast vote and the voting declaration status.
Selection of the right variables to remove requires a good understanding of the application domain; we assume that it is provided by the user.
Roughly speaking, the abstraction procedure takes the combined MAS graph $comb(G)$, computes an approximation of the reachable values for every $v\in V$, and processes the edges of $comb(G)$ by substituting the occurrences of $v$ at location $\ell$ with the values $u\in appr(v,\ell)$.
If $appr(v,\ell)$ overapproximates (resp.~underapproximates) the actual reachable values of $v$ at $\ell$, then the resulting model is a \emph{may} (resp.~\emph{must})-abstraction of $G$.

\para{Variable merge\short{ and scoping}.}
More generally, a subset of variables can be merged into a fresh variable by means of a user-defined mapping function.
For example, \texttt{mem\_sg} and \texttt{mem\_vt} can be merged into a boolean variable \texttt{valid} given by \texttt{(mem\_sg*mem\_vt>0)}, indicating the validity of the vote.

\extended{\para{Scope.}}
Additionally the user can specify the scope of abstraction, i.e., a subset of locations where the abstraction is applied.
\extended{
  , so that it only takes effect (and discards some details from the state) on a fragment of the system (when those details are no longer relevant/needed).
  This should facilitate in refining abstraction and help to obtain a conclusive result from verification of an abstract model.
For example, we could remove variable \texttt{mem\_dec}, which encodes delivery medium, at locations \texttt{has} and \texttt{end}, after the election package was collected.

  Depending on the model, property and user's creativity, one can come up with plenty other interesting variant of abstraction with this tool.
  For example, when there is a set of indistinguishable agents of the same type (say Voters) and property has to do with capabilities of one or few agents only, then we could also use a coarser abstraction for all except one or few selected.
}

\para{Abstraction on MAS templates.}
In some cases, approximation of variable domains on the combined MAS graph is computationally infeasible\extended{ due to the size of the graph}.
An alternative is to compute it directly on the MAS template by the right approximation of the synchronization edges.
On the down side, this sometimes results in largely suboptimal abstract models, i.e., ones more likely to produce inconclusive verification results.

\extended{
  procedure may be run on either combined MAS graph or just an agent template.\footnote{ A case of partially combined MAS graph can also be defined, but that goes beyond the interest of present discussion.
  } The former is expected to result with a more accurate approximation, but its computation also demands more resources (such as time and memory); the latter is less accurate, but is also less resource demanding.
  Unfortunately, the problem of state space explosion may occur even at the level of combined MAS graph representation.
  This becomes more evident on models with large number of agents.
  The intuition/heuristic is to choose a template-based approximation, when the template agent does not have many synchronous edges or those synchronisations are not expected to have much impact on target variables evaluation \YK{($\leftarrow$ way too informal?)}.
  For example, local domain for \texttt{mem\_vt} and \texttt{mem\_sg}, for which values are assigned locally (i.e., with no transitive reference to any global variable), could as well be approximated on the template instead; this
}

\begin{table}[t]\centering
    \resizebox{\columnwidth}{!}{

\small{
\begin{tabular}{|c||rr||rr||r|r||r|r|}
    \hline
    \multirow{2}{*}{\textbf{\#V}} & \multicolumn{2}{c||}{Concrete}      & \multicolumn{2}{c||}{Abstract (A1)}    & \multicolumn{2}{c||}{Abstract (A2)} & \multicolumn{2}{c|}{Abstract (A3)}                                                                                                              \\ \cline{2-9}
                                  & \multicolumn{1}{c|}{\textbf{\#St}}  & \multicolumn{1}{c||}{\textbf{t}}    & \multicolumn{1}{c|}{\textbf{\#St}} & \multicolumn{1}{c||}{\textbf{t}} & \multicolumn{1}{c|}{\textbf{\#St}} & \multicolumn{1}{c||}{\textbf{t}} & \multicolumn{1}{c|}{\textbf{\#St}} & \multicolumn{1}{c|}{\textbf{t}} \\ \hhline{|=::==::==::==::==|} 1                             & \multicolumn{1}{r|}{31}             & 0                                   & \multicolumn{1}{r|}{23}            & 0                               & 22                & 0      & 18                                 & 0                               \\ \hline
    2                             & \multicolumn{1}{r|}{529}            & 0.1                                 & \multicolumn{1}{r|}{217}           & 0.1                             & 214               & 0.1    & 120                                & 0.1                             \\ \hline
    3                             & \multicolumn{1}{r|}{10891}          & 0.1                                 & \multicolumn{1}{r|}{2203}          & 0.1                             & 2440              & 0.1    & 838                                & 0.1                             \\ \hline
    4                             & \multicolumn{1}{r|}{2.3e+5}         & 0.9                                 & \multicolumn{1}{r|}{22625}         & 1                               & 29938             & 0.1    & 5937                               & 0.1                             \\ \hline
    5                             & \multicolumn{1}{r|}{5.1e+6}         & 25                                  & \multicolumn{1}{r|}{2.3e+5}        & 1                               & 3.7e+5            & 1      & 42100                             & 0.6                             \\ \hline
6                             & \multicolumn{2}{c||}{memout} & \multicolumn{1}{r|}{2.3e+6}         & 20                                                                              & 4.9e+6            & 23     & 2.9e+5                           & 5                                                                    \\ \hline
7                             & \multicolumn{2}{c||}{memout} & \multicolumn{1}{r|}{2.2e+7}         & 304                                                                             & \multicolumn{2}{c||}{memout}                     & 2.0e+6                           & 33                                                                   \\ \hline
8                             & \multicolumn{2}{c||}{memout} & \multicolumn{2}{c||}{memout}                                                                                          & \multicolumn{2}{c||}{memout}                     & 1.4e+7                             & 357                                                                                                     \\ \hline
\end{tabular}
}
 }
    \caption{Verification of $\varphi_\mathit{bstuff}$ on models with 3 candidates. $\#V$ is the number of Voter instances. We report the model checking performance for the concrete model, followed by \emph{may}-models obtained by abstractions A1, A2, and A3}
    \label{tab:results-postal}
\end{table}

\section{Architecture}

The main components of the tool are: (1) local domain approximation and (2) generation of abstract model specifications.
Additionally, the tool allows to perform simple pre-processing and code analysis, and to store parameters in a configuration file.
Each component can be called from command line, possibly followed by a list of arguments:
\begin{itemize2}
\item \texttt{configure}: sets the parameters in the configuration file;
\item \texttt{unfold}: produces the combined MAS graph;
\item \texttt{approx}: computes an approximation of the local domain;\item \texttt{abstract}: generates an abstract model specification based on the provided approximation of local domain;
\item \texttt{info}: lists the variables, locations, and edges in the model.
\end{itemize2}

\extended{
  \para{Unfold.}
  Substitutes the constants with their values and converts the MAS graph from the input into the combined MAS graph.
}

\para{Local domain approximation.}
Takes a subset of variables $V$, a target template (`ext' for the combined MAS graph) and an abstraction type $\text{t}\in\set{\text{upper}, \text{lower}}$, and computes a $\text{t}$-approximation of the local domain over $V$.
The result is saved to a JSON file, where location identifiers are mapped to an array of evaluation vectors.
\extended{Note that the order of vector elements will correspond to an order of previously given variable identifiers.}

\para{Abstract model generation.}
Takes the mapping function with an upper-approximation (resp.~lower-approximation) of the local domain, and computes the corresponding may-abstraction (resp. must-abstraction).
The mapping function specifies the target agent name or template name, the scope of abstraction\extended{ (subset of location that should be affected)}, variables to be removed, and possibly a merge variable\extended{ (name, initial value and evaluation expression)}.
We assume that the input provided by the user is correct; some debugging might be added in the future.

\extended{\para{Implementation details.}
The tool is written in \texttt{node.js}, which parses XML model specifications compatible with \Uppaal.
The command line interface has been created with the help of the \texttt{yargs} library.
After the model is parsed, its structure is processed using \texttt{antlr4}.
The formal syntax is given by the EBNF grammar provided in file \texttt{yag.g4}.}

\section{Experimental Results}

We have evaluated the tool by means of experiments on two benchmarks: a simple postal voting scenario and gossip learning for social AI.
The model specifications are available for download with the tool.
The experiments have been performed in combination with \Uppaal v4.1.24 (32-bit) on a machine with \worklaptopcpu, running Ubuntu 22.04.
We report the results for \emph{may-abstractions}, typically more useful for universal branching-time properties.

\extended{
Due to Uppaal only supporting a subset of \CTLs formulae (with non-standard interpretation of \todo{AF} type of formulae), verification of which is performed on-the-fly \todo{add citation}, results will be mainly focused around may-abstraction and \todo{AG} formulae.\footnote{To show that the model does not satisfy an AG formula it suffices to find a witness for an equivalent EF formula and is often easy in practice (thus, use of must-abstraction may show little to no improvement in terms of verification time).
    It becomes even more so with use of on-the-fly verification technique as both generation of states and examination of model are done simultaneously.
    Nevertheless, it is worth noting that must-abstraction could be of immense importance when model is generated prior to its exploration (e.g., in case of \CTLs\ formulae with nested path quantifiers). }
}

\para{Postal voting.}
We use a scalable family of MAS graphs proposed in~\cite{Kim22postal-stast} to model a simplified postal voting system.
The system consists of \NVot\ Voters, voting for \NCand\ candidates, and a single Election Authority, and proceeds in four subsequent phases: collection of voting declarations, preparation and distribution of election packages, ballot casting, and tallying.
The verification concerns a variant of resistance to ballot stuffing, expressed by formula $\varphi_\mathit{bstuff}$:\\
\centerline{\texttt{A[](b\_recv<=ep\_sent \&\& ep\_sent<=NV)}}\\
where \texttt{b\_recv} and \texttt{ep\_sent} are variables storing the number of received ballots and sent election packages, respectively.
For the experiments, we try the following abstractions:
\begin{description2}
\item[A1:] removes variables \texttt{mem\_vt} and \texttt{mem\_sg} from the Voter template, i.e., the voter's memory of the cast vote and the voting declaration status;
\item[A2:] removes variables \texttt{mem\_dec} at Voter's locations \{\texttt{has}, \texttt{voted}\} and variable \texttt{dec\_recv} at Authority's location \{\texttt{coll\_vts}\}, i.e., the information about how the election package has been delivered;
\item[A3:] the combination of A1 and A2.
\end{description2}
The results in~\autoref{tab:results-postal} present the numbers of states in the global model generated during the verification, as well as the verification running times (in seconds), including the generation of abstract model specifications where applicable.
Formula $\varphi_\mathit{bstuff}$ is satisfied in all the reported instances; all three abstractions have been conclusive on it.

\begin{table}[t]\centering
  \resizebox{0.789\columnwidth}{!}{

\small{
    \begin{tabular}{|c||cr||r|c|r|}
    \hline
    \multirow{2}{*}{\textbf{\#Ag}} & \multicolumn{2}{c||}{Concrete}      & \multicolumn{3}{c|}{Abstract}    \\ \cline{2-6}
                  & \multicolumn{1}{c|}{\textbf{\#St}} & \multicolumn{1}{c||}{\textbf{t}} & \multicolumn{1}{c|}{\textbf{\#St}} & \multicolumn{1}{c|}{\textbf{Reduct}} & \multicolumn{1}{c|}{\textbf{t}} \\ \hhline{|=::==::===|}2             & \multicolumn{1}{r|}{165}           & 0                               & 38                                    & 76.97                                                 & 0                                  \\ \hline
    3             & \multicolumn{1}{r|}{8917}          & 0.1                             & 555                                   & 93.78                                                 & 0                                  \\ \hline
    4             & \multicolumn{1}{r|}{4.6e+5}        & 1.5                             & 10247                                  & 97.77                                                 & 0.1                                \\ \hline
    5             & \multicolumn{1}{r|}{2.1e+7}        & 123                             & 1.5e+5                                  & 99.29                                                 & 1.2                                \\ \hline
    6             & \multicolumn{2}{c||}{memout}                                          & 2.8e+6                                & \multicolumn{1}{c|}{--}                                   & 42                                \\ \hline
    7             & \multicolumn{2}{c||}{memout}                                          & 4.1e+7                                & \multicolumn{1}{c|}{--}                                   & 682                                  \\ \hline
    8             & \multicolumn{2}{c||}{memout}                                          & \multicolumn{3}{c|}{memout}            \\ \hline
    \end{tabular}
}

 }
  \caption{Verification of $\varphi_\mathit{compr}$ on models of social AI. $\#Ag$ is the number of agents. ``Reduct'' shows the level of reduction of the state space (in \%) }
  \label{tab:results-sai}
\end{table}

\para{Social AI.}
The second series of experiments uses the specifications of gossip learning for social AI~\cite{Heaven13gossip,Hegedus21gossip-vs-federated-learning}, proposed in~\cite{Kurpiewski23SAI}.
The system consists of a ring network of AI agents, acting in three phases: data gathering, learning\extended{ (based on the previously collected data)}, and sharing of knowledge.
The goal of the agents is to collectively reach knowledge of quality $mqual\ge 2$. The system includes also an attacker who can impersonate any agent and fake its quality level.
\short{The model specification given as asynchronous MAS~\cite{Jamroga20POR-JAIR} and coded in the input language of the STV model checker~\cite{Kurpiewski21stv-demo} was manually translated into the input language of \Uppaal.}
\extended{The model specifications are given as asynchronous MAS~\cite{Jamroga20POR-JAIR\extended{,Jamroga20paradoxes-tr}}, and coded in the input language of the STV model checker~\cite{Kurpiewski21stv-demo}, that shares many similarities with the input language of \Uppaal.
}
\extended{After a straightforward manual translation to \Uppaal}\short{Afterwards}, we hardcoded the attacker's strategy to always share the lowest quality model, and verified formula $\varphi_\mathit{compr}$:\\
  \centerline{\small\texttt{A[](exists(i:int[1,NA])(impersonated!=i \&\& }}\\
  \centerline{\small\texttt{(!AI(i).wait || AI(i).mqual<2))). }}\\
$\varphi_\mathit{compr}$ says that, on all execution paths, at least one AI agent is compromised.
The model checking performance is shown in~\autoref{tab:results-sai}.
We have been able to conduct verification for concrete models with up to 5 agents (4 honest AI and 1 attacker), and up to 7 agents after applying a \emph{may}-abstraction that discards all variables except for $mqual$ in the AI template.
\extended{
  It took less than 1s to do a template-based over-approximation of a local domain followed by a generation of an abstract model.\footnote{
    Used commands can be found in the demo video and on the github. }
}

\extended{
  and approximate the strategic abilities of a coalition of honest agents by themselves and in presence of an active impersonator type of attacker.\footnote{Commonly, such kind of properties are formalized using an alternating-time temporal logic \atl (and its variants) \cite{Alur02ATL} and extends \CTLs with strategic modalities, which are, unfortunately, not supported by Uppaal.
    Nevertheless, an \atl formula of the form $\coop{A}{\Diamond \varphi}$ can still be loosely approximated by $\Apath \Diamond \varphi$ and $\Epath \Diamond \varphi$ as an over and under approximations respectively. }
}

\section{Conclusions}

We propose a tool for practical model reductions in multi-agent systems.
The tool addresses state-space explosion by removal of selected variables from the model while preserving the truth of \ACTL formulas.
The experiments show significant gains in terms of verification time as well as memory, with minimal time used by the abstraction procedure. \extended{The procedure directly modifies the modular specification of the system, without generating the global model at all.
Moreover, its output is open for further scrutiny and modifications by the user.
}

In the future, we plan to extend our tool to abstractions preserving temporal-epistemic and strategic properties in combination with the MCMAS and STV \mbox{model checkers~\cite{Lomuscio17mcmas,Kurpiewski21stv-demo}.}

\section*{Acknowledgments}
The work was supported by NCBR Poland and FNR Luxembourg under the PolLux/FNR-CORE projects STV (POLLUX-VII/1/2019) and SpaceVote (POLLUX-XI/14/SpaceVote/2023), as well as the CHIST-ERA grant CHIST-ERA-19-XAI-010 by NCN Poland (2020/02/Y/ST6/00064).

\bibliographystyle{named}

\end{document}